\documentclass[prb,aps,superscriptaddress,twocolumn]{revtex4}

\usepackage{graphicx}
\usepackage{dcolumn}
\usepackage{bm}
\usepackage{amssymb}

\begin{document}


\title{Quantum 120$^\circ$ model on pyrochlore lattice:
orbital ordering in MnV$_2$O$_4$}

\author{Gia-Wei Chern}
\affiliation{Department of Physics, University of Wisconsin,
Madison, Wisconsin 53706, USA}

\author{Natalia Perkins}
\affiliation{Department of Physics, University of Wisconsin,
Madison, Wisconsin 53706, USA}

\author{Zhihao Hao}
\affiliation{Department of Physics and Astronomy, Johns Hopkins
University, Baltimore, Maryland 21218, USA}
\date{\today}

\begin{abstract}
We present an analytical model of orbital ordering in vanadium
spinel MnV$_2$O$_4$. The model is based on recent first-principles
calculation indicating a strong trigonal distortion at the vanadium
sites of this compound [Phys. Rev. Lett. {\bf 102}, 216405 (2009)].
At the single-ion level, the trigonal crystal field leaves a doubly
degenerate atomic ground state and breaks the approximate rotational
symmetry of $t_{2g}$ orbitals. We find that the effective
interaction between the low-energy doublets is described by a
quantum antiferromagnetic 120$^\circ$ model on the pyrochlore
lattice. We obtain the classical ground state and show its stability
against quantum fluctuations. The corresponding orbital order
consisting of two inequivalent orbital chains is consistent with the
experimentally observed tetragonal symmetry. A periodic modulation
of electron density function along orbital chains is shown to arise
from the staggering of local trigonal axes. In the presence of
orbital order, single-ion spin anisotropy arising from relativistic
spin-orbit interaction stabilizes the experimentally observed
orthogonal magnetic structure.
\end{abstract}

\maketitle

\section{Introduction}

Geometrically frustrated magnets with orbital degeneracy exhibit a
variety of complex ground states with unusual magnetic and orbital
orders. \cite{pen97,radaelli02,schmidt04,horibe06}  Not only do
these spin-orbital models deepen our understanding of systems with
competing degrees of freedom, they also describe the low-energy
physics of several transition-metal compounds. Of particular
interest is spin-orbital model on three-dimensional pyrochlore
lattice, \cite{dimatteo04,khomskii05} where geometrical frustration
between nearest-neighbor spins leads to a macroscopic degeneracy in
the classical ground state.\cite{moessner98} Experimentally, a
signature of strong frustration is the occurrence of a magnetic
phase transition well below the Curie-Weiss temperature. For systems
with degenerate orbitals, magnetic frustration is partially relieved
in the presence of a long-range orbital order, which is usually
accompanied by a simultaneous structural distortion due to
Jahn-Teller effect. The resulting magnetic order depends critically
on the orbital configuration and the details of spin-orbital
interactions.

Recently, much attention has been focused on vanadium spinels
$A$V$_2$O$_4$, where vanadium ions occupying the structural
$B$-sites form a pyrochlore lattice.
\cite{mamiya97,reehuis03,onoda03,lee04,adachi05,suzuki07,garlea08,chung08}
The two $d$ electrons of V$^{3+}$ ion have a total spin $S=1$ and
occupy two out of three $t_{2g}$ orbitals. Thermodynamically,
vanadium spinels with a non-magnetic $A$-site ion ($A =$ Zn, Cd, and
Mg) exhibit similar behavior: a structural transition which lowers
the crystal symmetry from cubic $Fd\bar 3m$ to tetragonal $I4_1/amd$
is followed by a magnetic ordering at a lower temperature.
\cite{mamiya97,reehuis03,onoda03} Assuming that $xy$ orbital is
occupied at all sites due to flattened VO$_6$ octahedra,
minimization of a Kugel-Khomskii type Hamiltonian on pyrochlore
lattice gives rise to a staggered ordering of the remaining $yz$ and
$zx$ orbitals (so-called $A$-type ordering). \cite{tsunetsugu03} The
resulting symmetry $I4_1/a$, however, is incompatible with
experimental observations.

On the other hand, assuming a large relativistic spin-orbit
coupling, a ferro-orbital order in which one electron occupies the
low-energy $xy$ orbital, whereas the other one is in states
$|yz\rangle \pm i|zx \rangle$ has been proposed in
Ref.~\onlinecite{ot04}. The occurrence of complex orbitals imply a
nonzero orbital angular momentum. This model successfully explains
important experimental results: the uniform occupation of orbitals
is compatible with space group $I4_1/amd$; an ordering of orbital
moment opposite to local spin direction is also consistent with the
observed reduced vanadium moment. A ground state with complex
orbitals has also been confirmed by mean-field \cite{dimatteo05} and
{\em ab initio} calculations. \cite{maitra07,pardo08}

Interest in antiferro-orbital order is rekindled by a recent
experimental characterization of another vanadium spinel
MnV$_2$O$_4$, \cite{adachi05,suzuki07,garlea08,chung08} where the
$A$-site Mn$^{2+}$ ion is in a $3d^5$ high spin configuration ($S =
5/2$). In contrast to other vanadium spinels, MnV$_2$O$_4$ first
undergoes a magnetic transition at $T_F = 56$ K into a collinear
ferr{\em i}magnetic phase with Mn and V moments aligned antiparallel
to each other. At a slightly lower temperature $T_S = 53$ K, a
structural distortion lowering the crystal symmetry to tetragonal
$I4_1/a$ is accompanied by an ordering of the transverse components
of V spins. The ground-state orbital configuration is suggested to
be the $A$-type antiferro-orbital order. \cite{garlea08} Also
contrary to collinear magnetic order in other vanadium spinels, a
peculiar non-collinear order with transverse component of vanadium
spins forming an orthogonal structure in the $ab$ plane was observed
in MnV$_2$O$_4$. \cite{garlea08,chung08}

Recently we have demonstrated the stability of orthogonal magnetic
structure in the limit of strong relativistic spin-orbit coupling.
\cite{chern08,chern09} However, findings from first-principles
calculation indicate a significant trigonal distortion at the
vanadium sites of MnV$_2$O$_4$, \cite{sarkar09} whose effect is yet
to be understood. The same authors find an orbital order consisting
of two inequivalent orbital chains similar to the $A$-type order.
More importantly, they observe an additional modulation of electron
density profile {\em within} each orbital chain: the orbitals rotate
alternatively by about 45$^\circ$ along the chain. This complex
orbital pattern is also supported by a recent NMR measurement.
\cite{baek09}

Based on these observations, we present an analytical model of
spinel MnV$_2$O$_4$ assuming that the $t_{2g}$ orbitals is split
into a singlet and a doublet by a strong trigonal crystal field. As
one electron occupies the low-energy $a_{1g}$ orbital, a double
degeneracy remains for the other electron. After introducing a
pseudospin-1/2 to describe the doubly degenerate atomic ground
state, we find that their effective interaction is governed by a
highly anisotropic quantum 120$^\circ$ Hamiltonian.
\cite{nussinov04,biskup05} By treating quantum fluctuations using
the semiclassical framework, the classical ground state of the
120$^\circ$ model is shown to be stable against quantum
fluctuations. Orbital ordering and lattice distortion derived from
the classical ground states are consistent with the experiments. We
also shown that an alternatively rotated orbital basis due to the
staggered trigonal axes  explains the periodic density-modulation
observed in {\em ab initio} calculations. Moreover, since the very
presence of trigonal distortion breaks the (approximate) rotational
symmetry of $t_{2g}$ orbitals, orthogonal magnetic structure thus
comes naturally from spin-orbit interaction and the staggering of
trigonal axes.

The rest of the paper is organized as follows. Sec.~\ref{sec:m120}
discusses the effective orbital 120$^\circ$ model on pyrochlore
lattice and its semiclassical ground states. The corresponding
orbital order and lattice distortion are discussed in
Sec.~\ref{sec:o-order}. The modulation of electron density function
along orbital chain is addressed in Sec.~\ref{sec:modulation}. The
details of magnetic structure is presented in
Sec.~\ref{sec:m-order}. And finally Sec.~\ref{sec:conclusion}
presents a conclusion.

\section{120$^\circ$ model}
\label{sec:m120}

The site symmetry of vanadium ions in most vanadates is dominated by
a cubic crystal field. Nonetheless, splitting of $t_{2g}$ triplet
due to an additional trigonal distortion is known to play an
important role is some cases. Most notably, stabilization of the
unusual magnetic structure in the insulating phase of V$_2$O$_3$ can
only be understood when the trigonal splitting is properly taken
into account. \cite{castellani,mila00,perkins0209} The effects of
trigonal distortions in vanadium spinels vary from one compound to
another. For example, the trigonal splitting of $t_{2g}$ levels is
essential to the understanding of heavy fermion behavior in metallic
LiV$_2$O$_4$. \cite{anisimov99} On the other hand, it seems to have
a negligible effect in another well-studied spinel ZnV$_2$O$_4$.
\cite{maitra07}

Recently, experiment \cite{chung08} and {\em ab~initio} calculation
\cite{sarkar09} both indicate a strong trigonal distortion at the
vanadium sites of MnV$_2$O$_4$. In order to understand its effects
at least qualitatively and to make analytical calculations
tractable, we consider the limit of a dominating trigonal crystal
field in this paper. As discussed in the introduction, the trigonal
distortion still leaves a doubly degenerate atomic ground state. The
possible long-range order of these localized doublets is
investigated using the effective Hamiltonian approach. By studying
the ground state of the effective model, we discuss its implications
for orbital and magnetic ordering in MnV$_2$O$_4$.

\begin{figure}
\centering
\includegraphics[width=0.85\columnwidth]{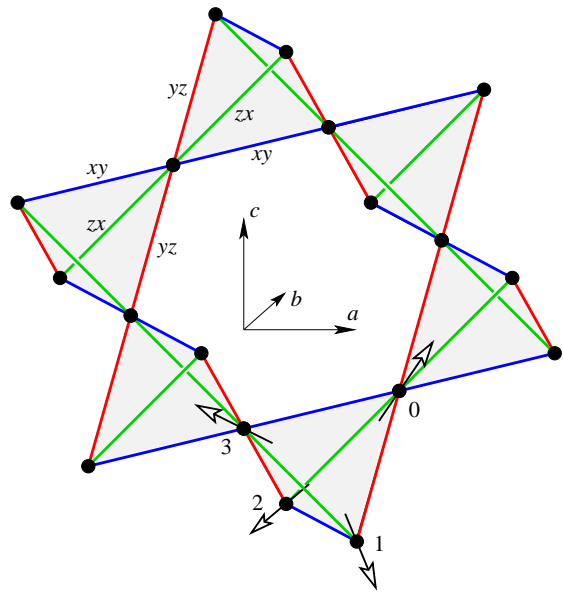}
\caption{\label{fig:pyrochlore} Pyrochlore lattice. The numbers 0--3
denotes the four sublattices of pyrochlore lattice; the arrows
indicate the local $C_3$ axis of the respective sublattices.
Explicitly, they are $\hat{\bm\nu}_0 = [111]$, $\hat{\bm\nu}_1 =
[1\bar 1\bar 1]$, $\hat{\bm\nu}_2 = [\bar 1 1 \bar 1]$, and
$\hat{\bm\nu}_3 = [\bar 1\bar 1 1]$. In 120$^\circ$-model Eq.
(\ref{eq:H120}), orbital interactions on red, green, and blue bonds
are characterized by vectors $\hat\mathbf n_{yz}$, $\hat\mathbf
n_{zx}$, and $\hat\mathbf n_{xy}$, respectively.}
\end{figure}

In the presence of a trigonal distortion, the crystal field of
reduced site symmetry (from cubic $O_h$ to $D_{3d}$) splits $t_{2g}$
orbitals into a singlet and a doublet separated by an energy gap
$\Delta$. The $C_3$ symmetry axis of $D_{3d}$ group is parallel to
the local $\langle 111 \rangle$ direction of the ion (Fig.
\ref{fig:pyrochlore}). The $a_{1g}$ singlet is the symmetric linear
combination of $t_{2g}$ orbitals under $C_3$ rotation
\begin{eqnarray}
    |a_{1g}\rangle = \nu_x |yz\rangle
    + \nu_y |zx\rangle + \nu_z |xy\rangle,
\end{eqnarray}
where $\hat{\bm\nu}=(\nu_x, \nu_y,\nu_z)$ is a unit vector parallel
to the local trigonal axis (Fig.~\ref{fig:pyrochlore}). We use the
following chiral basis for the $e_g$ doublet
\begin{eqnarray}
    \label{eq:eg}
    \begin{array}{c}
    |e_g^+\rangle  = \nu_x e^{-i\omega} |yz\rangle +
    \nu_y e^{+i\omega} |zx \rangle + \nu_z |xy\rangle, \\
    |e_g^-\rangle  = \nu_x e^{+i\omega} |yz\rangle +
    \nu_y e^{-i\omega} |zx \rangle + \nu_z |xy\rangle,
    \end{array}
\end{eqnarray}
where $\omega = 2\pi/3$. A complete basis for V$^{3+}$ ion with
$3d^2$ configuration is given by $|a_{1g} e^+_g \rangle$, $|a_{1g}
e^-_g\rangle$, and $|e^+_g e^-_g\rangle$. Here the two-electron
state is defined as the antisymmetric sum of individual one-electron
states, i.e. $|\alpha\beta\rangle \equiv
\bigl(|\alpha\rangle|\beta\rangle -
|\beta\rangle|\alpha\rangle\bigr)/\sqrt{2}$. Since the $a_{1g}$
singlet has the lowest energy, the atomic ground state is doubly
degenerate. To describe the low-energy doublet manifold, we
introduce a pseudospin-$\frac{1}{2}$ operator $\bm \tau$ such that
$|\tau_z = \pm 1\rangle$ are identified with $|a_{1g}
e_g^\pm\rangle$, respectively.

\begin{figure}
\centering
\includegraphics[width=0.52\columnwidth]{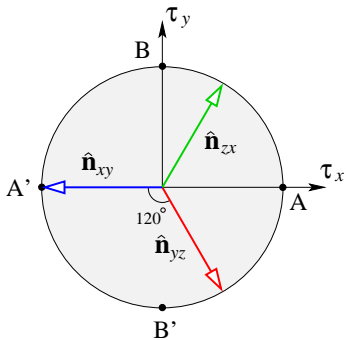}
\caption{\label{fig:circle} Unit circle in the $(\tau_x, \tau_y)$
plane. Pseudospins $\bm\tau_i$ participate in the 120$^\circ$ interaction
(\ref{eq:H120}) only through their projections onto the
three unit vectors $\hat\mathbf n_{yz}$, $\hat\mathbf n_{zx}$,
and $\hat\mathbf n_{xy}$.}
\end{figure}

In vanadium spinels, the superexchange (SE) interaction with a
$90^\circ$ angle between vanadium-oxygen bonds is dominated by
direct exchange which involves electron hopping of the $dd\sigma$
type: \cite{tsunetsugu03,dimatteo05}
\begin{eqnarray}
    \label{eq:SE}
    \!\!& &H_{\rm SE} = -\sum_{\langle ij \rangle}\Bigl\{
    J_2\bigl(1-\mathbf S_i\!\cdot\!\mathbf S_j)\, P_{\alpha,i}
    P_{\alpha,j}
    \\& &\quad +
    \bigl(J_0\,\mathbf S_i\!\cdot\!\mathbf S_j +
    J_1\bigr)\bigl[P_{\alpha,i}(1-P_{\alpha,j})
    + (1-P_{\alpha,i}) P_{\alpha,j}\bigr]\Bigr\}. \nonumber
\end{eqnarray}
The various exchange constants are $J_0 = J\eta/(1-3\eta)$, $J_1 = J
(1-\eta)/(1-3\eta)$, and $J_2 = J (1+\eta)/(1+2\eta)$, where $J =
t_{dd\sigma}^2/U$ sets the overall energy scale and $\eta =
J_H/U\approx 0.11$ denotes the ratio of Hund's exchange to on-site
Coulomb repulsion. The subscript $\alpha \equiv \alpha(ij)$ of the
projection operators specifies the type of orbitals in which
electron hopping is possible between sites $i$ and $j$, e.g.
$\alpha(ij) = xy$ for nearest-neighbor bonds on $\langle 110
\rangle$ and $\langle 1\bar 1 0\rangle$ chains (blue bonds in Fig.
\ref{fig:pyrochlore}).

An interesting feature of Hamiltonian (\ref{eq:SE}) is the static
Potts-like orbital interactions which  depend  only on orbital
projection operators  $P_{xy}$, $P_{yz}$, and $P_{zx}$. Restricted
to the doublet manifold $|\tau_z = \pm1 \rangle$, they become
\begin{eqnarray}
    \label{eq:P}
    P_\alpha = \frac{2}{3} - \frac{1}{3}\,\bm \tau
    \cdot \hat\mathbf n_\alpha,
\end{eqnarray}
where the three unit vectors are (Fig.~\ref{fig:circle}):
\begin{eqnarray}
    \label{eq:unit-n}
    \hat\mathbf n_{yz} = \frac{1}{2}\hat\mathbf x -
    \frac{\sqrt{3}}{2}\hat\mathbf y, \quad
    \hat\mathbf n_{zx} = \frac{1}{2}\hat\mathbf x +
    \frac{\sqrt{3}}{2}\hat\mathbf y, \quad
    \hat\mathbf n_{xy} = -\hat\mathbf x.\quad
\end{eqnarray}
A remark is now in order: the restricted Hilbert space $|\pm\rangle$
already precludes descriptions of, e.g.~the $A$-type
antiferro-orbital order consisting of alternating $|xy,yz\rangle$
and $|xy,zx\rangle$ states. \cite{tsunetsugu03} A more general
approach is to introduce a pseudospin-1 formulation similar to the
one used in Ref.~\onlinecite{perkins0209}; the trigonal splitting is
then modeled by a spin anisotropy term. However, the resulting
effective Hamiltonian is quite complicated and analytical
calculations are difficult. In this paper, we choose the simplified
pseudospin-$\frac{1}{2}$ formulation to explore the essential
features of a large trigonal distortion and use perturbation method
to examine the effect of excited state $|e^+_g e^-_g\rangle$.

The effective Hamiltonian $H_{\rm eff}$ of pseudospins $\bm \tau$
can be obtained by projecting SE Hamiltonian onto the doublet
manifold, or equivalently by substituting the projection operators
(\ref{eq:P}) into Eq. (\ref{eq:SE}). In order to understand the
intrinsic properties of orbital interaction, we first examine
$H_{\rm eff}$ above the structural transition $T_S$, where
nearest-neighbor spin correlations $\langle \mathbf S_i\cdot\mathbf
S_j\rangle$ are isotropic. For example, below the Curie-Weiss
temperature, the magnet is in a strongly correlated liquid-like
state, \cite{moessner98} the constraint of zero total spin $\mathbf
S_{\boxtimes} = 0$ on every tetrahedron gives a spin correlation
$\langle \mathbf S_i\cdot\mathbf S_j \rangle = -S^2/3$. In the
collinear ferrimagnetic phase, a partial ferromagnetic order
$\mathbf M$ antiparallel to the Mn moments is induced by the
antiferromagnetic Mn-V exchange; the spin correlation becomes
$\langle \mathbf S_i\cdot\mathbf S_j\rangle = -S^2/3 + M^2$. In all
cases, the effective orbital interaction has the following
anisotropic form:
\begin{eqnarray}
    \label{eq:H120}
    H_{\rm eff} = J_{\tau}\sum_{\langle ij \rangle}
    (\bm \tau_i\cdot\hat\mathbf n_{\alpha})
    (\bm\tau_j\cdot\hat\mathbf n_{\alpha}),
\end{eqnarray}
where $J_{\tau} = \frac{1}{9}[2J_1 - J_2 + \langle \mathbf
S_i\cdot\mathbf S_j\rangle (J_2 + 2 J_0)] > 0$
 and $\alpha =
\alpha(ij) = yz$, $zx$, and $xy$ depending on the orientation of the
nearest-neighbor bond (Fig.~\ref{fig:pyrochlore}). The effective
Hamiltonian has a form of the so-called 120$^\circ$ model,
\cite{nussinov04,biskup05} which was first introduced as an
effective model for perovskite $e_g$ orbital systems.
\cite{brink99,khomskii03,note} Recently, the same model was found to
describe the insulating phase of $p$-band fermions in optical
lattices. \cite{zhao08,wu08} The 120$^\circ$ model is closely
related to the well-known quantum compass model. \cite{kk} A common
feature shared by these highly anisotropic spin models is the
competition between bonds along different directions. For compass
and 120$^\circ$ models on bipartite lattices, a macroscopic
degeneracy of the ground state results from the discrete gauge-like
sliding symmetries. \cite{nussinov05} Remarkably, as discussed in
more detail below, such extensive degeneracy is absent in the
classical 120$^\circ$ model on non-bipartite pyrochlore lattice.

We first discuss the origin of the macroscopic degeneracy in the
cubic lattice 120$^\circ$ model.  The three unit vectors
$\hat\mathbf n_{\alpha}$ in Eq.~(\ref{eq:unit-n}) are associated
with nearest-neighbor bonds along $x$, $y$, and $z$ directions,
respectively. \cite{khomskii03} The bipartite nature of the cubic
lattice allows us to transform an antiferromagnetic coupling to a
ferromagnetic one through a $\pi$-rotation about $\tau_z$ axis on
one sublattice. An unusual property of the classical model is the
appearance of planar gauge-like symmetries in addition to global
spin rotations. \cite{nussinov04,biskup05} A huge ground-state
degeneracy thus results from the gauge-like $Z_2$ transformations.
As demonstrated in Ref.~\onlinecite{biskup05}, starting from a state
of uniform spins, which is a ground state of the ferromagnetic
model, another inequivalent ground state can be obtained by rotating
all spins on a randomly chosen $xy$ plane by an angle $\pi$ about
$\tau_x$ axis. Nonetheless, long-range order arises via the
order-by-disorder mechanism which in general favors collinear
(uniform) spin configurations. \cite{nussinov04}

Interestingly, the above-mentioned gauge-like symmetry is absent for
120$^\circ$ model on pyrochlore lattice. This is because a
prerequisite for the $Z_2$ gauge-like transformation is the
existence of a subset $\mathcal{C}$ of lattice sites (e.g., planes
or chains) such that pseudospins belonging to the subset are
connected to each other by, say, either $yz$ or $zx$ bonds, while
interaction of pseudospin $i\in \mathcal{C}$ with its neighbor
$j\notin \mathcal{C}$ is of the $xy$ type exclusively. It could be
easily checked that such a subset cannot be found in pyrochlore
lattice. However, orbital interactions are still frustrated simply
due to geometry: antiferromagnetic pseudospin interaction cannot be
satisfied on all nearest-neighbor bonds simultaneously. \cite{note1}

\begin{figure} [t]
\centering
\includegraphics[width=0.95\columnwidth]{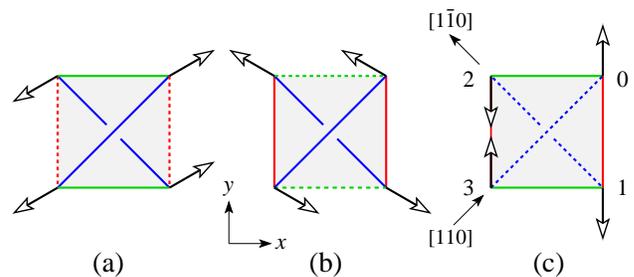}
\caption{\label{fig:gs} Ground states of the orbital
120$^\circ$-model. The other three ground states are related to the
above ones by a $C_2$ rotation of pseudospins about the $z$ axis.
Orbital interactions on red, green, and blue bonds are characterized
by vector $\mathbf n_{yz}$, $\mathbf n_{zx}$, and $\mathbf n_{xy}$,
respectively. The frustrated bonds (parallel pseudospins) are
indicated by dashed lines. Nonzero order parameter characterizing
the ground states are: (a) $\mathbf l_{yz} =
-\frac{\sqrt{3}}{2}\hat\mathbf x - \frac{1}{2}\hat\mathbf y$, (b)
$\mathbf l_{zx} = -\frac{\sqrt{3}}{2}\hat\mathbf x +
\frac{1}{2}\hat\mathbf y$, and (c) $\mathbf l_{xy} =  \hat\mathbf
y$.}
\end{figure}

Despite being geometrically frustrated, the strong anisotropy of
120$^\circ$ interaction significantly reduces the number of
degenerate ground states. To see this, we note that the energy of a
single bond is minimized classically by a pair of pseodospins
pointing toward $\pm \hat\mathbf n_{\alpha}$, respectively, where
$\hat\mathbf n_{\alpha}$ is the unit vector characterizing the
anisotropic interaction of the bond. However, such absolute minimum
can not be attained at every nearest-neighbor bonds due to
geometrical frustration. Even worse, pseudospin correlation on some
bonds is frustrated, i.e. $\bm\tau_i\cdot\bm\tau_j > 0$. In order to
minimize the energy cost, frustrated pseudospins thus tend to align
themselves perpendicular to $\hat \mathbf n_\alpha$. Through both
analytical calculation and Monte-Carlo simulations, we find that
collinear states with pseudospins perpendicular to either one of the
three $\hat \mathbf n_\alpha$ are the classical ground states
(Fig.~\ref{fig:gs}). The total degeneracy is six due to an
additional $C_2$-rotation about $\tau_z$ axis. This is in stark
contrast to the macroscopic ground-state degeneracy of classical
Heisenberg spins on the pyrochlore lattice. \cite{moessner98}

The three inequivalent ground states shown in Fig.~\ref{fig:gs} are
characterized by the locations of frustrated bonds. More
specifically, we introduce three staggered order parameters
\cite{chern06}
\begin{eqnarray}
    \mathbf l_{yz} &=& (\bm\tau_0 + \bm\tau_1 - \bm\tau_2 -
    \bm\tau_3)/4, \nonumber \\
    \mathbf l_{zx} &=& (\bm\tau_0 - \bm\tau_1 + \bm\tau_2 -
    \bm\tau_3)/4, \\
    \mathbf l_{xy} &=& (\bm\tau_0 - \bm\tau_1 - \bm\tau_2 +
    \bm\tau_3)/4 \nonumber
\end{eqnarray}
to describe the orbital order. Here $\bm\tau_i$ denotes pseudospin
average on $i$th sublattice. These order parameters measure the
difference of orbital configuration on bonds of the same type. For
example, a nonzero $\mathbf l_{xy}$ indicates an antiferro-orbital
order across the two $xy$ bonds of a tetrahedron [see
Fig.~\ref{fig:gs}(c)]. The nonzero order parameter characterizing
the collinear states of Fig.~\ref{fig:gs} are: (a) $\mathbf l_{yz} =
-\frac{\sqrt{3}}{2}\hat\mathbf x - \frac{1}{2}\hat\mathbf y$, (b)
$\mathbf l_{zx} = -\frac{\sqrt{3}}{2}\hat\mathbf x +
\frac{1}{2}\hat\mathbf y$, and (c) $\mathbf l_{xy} = \hat\mathbf y$.
This should be contrasted with the continuously degenerate collinear
ground states in Heisenberg model, i.e. $\mathbf l_{\alpha} =
\hat\mathbf e$ with $\hat\mathbf e$ being an arbitrary unit vector.
\cite{chern06}

The stability of classical ground states in the presence of quantum
fluctuations is investigated using the semiclassical
Holstein-Primakoff transformation. We find that the anisotropy of
the orbital exchange leads to a gapped quasiparticle spectrum in the
whole Brillouin zone. At the harmonic level, quantum fluctuations
around collinear ground states are shown to give a negligible
correction to the sublattice `magnetization' $\langle \bm\tau_i
\rangle$ (about 4\%), indicating the stability of the classical
ground states. A detailed account of the semiclassical calculation
is presented in Appendix \ref{sec:semi}.

Before closing this section, we remark that phonon-mediated orbital
exchange in spinels also has the form of 120$^\circ$ interaction
with an effective exchange $J_\tau \propto g^2/k_{F_{1g}}$, where
$g$ is a Jahn-Teller coupling constant and $k_{F_{1g}}$ is the
elastic constant of $F_{1g}$ phonons. \cite{chern08} Detailed
derivation is presented in Appendix \ref{sec:phonon}. In fact,
noting that $(\tau_x, \tau_y)$ forms a doublet irreducible
representation of $D_{3d}$ group, the 120$^\circ$ type interaction
in Eq.~(\ref{eq:H120}) is the only anisotropic pseudospin
interaction allowed by lattice symmetry. In this perspective, we
shall regard $J_{\tau}$ as an effective model parameter in the
following discussion.

\section{orbital order and lattice distortion}
\label{sec:o-order}

We now discuss the orbital order and lattice distortions
corresponding to the semiclassical ground states. Specifically, we
shall focus on the collinear state characterized by $\mathbf l_{xy}
= + \hat\mathbf y$ [Fig.~\ref{fig:gs}(c)]. Since
$P_{yz}+P_{zx}+P_{xy} = 2$, orbital orders are essentially described
by linear combinations
\begin{eqnarray}
    \label{eq:Pd}
    \begin{array}{c}
    P_1 = P_{yz} + P_{zx} - 2 P_{xy} = -\tau_x,   \\
    P_2 = \sqrt{3}(P_{zx} - P_{yz}) = -\tau_y,
    \end{array}
\end{eqnarray}
which transform as a doublet irreducible representation under
symmetry group $D_{3d}$. Since pseudospins pointing along $\pm y$
directions are sitting on $[110]$ and $[1\bar 10]$ chains,
respectively, the ground state shown in Fig.~\ref{fig:gs}(c)
consists of two distinct orbital chains characterized by $P_1 = 0$,
$P_2 = \mp 1$, respectively. The staggered part $P_2$ of the orbital
order comes from the occupation difference between $yz$ and $zx$
orbitals. The uniform part given by $P_{yz} = P_{zx} = P_{xy} = 2/3$
indicates that the three orbitals are equally occupied on average.

Due to Jahn-Teller effect, a long-range orbital order also implies a
lattice distortion in the ground state, which  is indeed observed in
MnV$_2$O$_4$ below $T_S = 53$ K. On symmetry ground, the coupling
between orbital doublet $|\tau_z = \pm1\rangle$ and distortions of
the surrounding VO$_6$ octahedron has the form:
\begin{eqnarray}
    \label{eq:jt}
    V_{\rm JT} = -g(\delta_1 \tau_x + \delta_2 \tau_y),
\end{eqnarray}
where $(\delta_1, \delta_2)$ are coordinates of normal modes
transforming as an $e_g$ representation of group $D_{3d}$. The two
symmetry-breaking modes can be thought of as analogous to the
tetragonal and orthorhombic distortions in a cubic VO$_6$
octahedron. The energy cost associated with the distortion is
$\frac{k}{2}\bigl(\delta_1^2 + \delta_2^2)$, where $k$ is an
effective elastic constant. For orbital order characterized by
$\mathbf l_{xy} = + \hat\mathbf y$, minimization with respect to
phonons yields distortions described by $\delta_1 = 0$ and $\delta_2
= \pm g/k$ on the two inequivalent orbital chains: octahedra on
$[110]$ and $[1\bar 10]$ chains are elongated along the $x$ and $y$
axes, respectively. The overall  distortion preserves the tetragonal
symmetry (lattice constants $a=b>c$) and is consistent with the
observed space group $I4_1/a$. \cite{garlea08}

It is interesting to note that the staggered orthorhombic
distortions of VO$_6$ octahedra actually correspond to a softened
$\mathbf q=0$ lattice phonons with $F_{1g}$ symmetry.
\cite{chern08,himmrich:91} This is consistent with the fact that, by
integrating out $F_{1g}$ phonons, orbital Jahn-Teller coupling gives
the same 120$^\circ$ pseudospin interaction
(Appendix~\ref{sec:phonon}). In this respect, orbital ordering and
structural transition in MnV$_2$O$_4$ can also be viewed as
softening of $F_{1g}$ phonons due to cooperative Jahn-Teller effect.

Despite the similarities between the antiferro-orbital order of
120$^\circ$ model and the $A$-type order proposed in
Ref.~\onlinecite{tsunetsugu03}, inclusion of spin-orbit interaction
illustrates an important difference between the two cases. As we
shall discuss later, spin-orbit coupling gives rise to an orthogonal
magnetic order in 120$^\circ$ model, which is in stark contrast to
collinear spins in the case of $A$-type orbital order.

\section{Modulation of electron density function}
\label{sec:modulation}

The staggering of trigonal axes along orbital chains also results in
a periodic variation of electron density distributions, despite the
orbital occupation numbers are invariant within the chain. This is
because the actual orbital wavefunction corresponding to $\bm\tau =
\pm \hat\mathbf y$ also depends on the local $C_3$ axis:
\begin{eqnarray}
    \begin{array}{l}
    |\tau_y = +1 \rangle = \sqrt{2}\bigl(\nu_x\cos\xi\, |X\rangle
    + \nu_y\sin\xi\, |Y\rangle\bigr)
    + \nu_z |Z\rangle,  \\
    |\tau_y = -1 \rangle = \sqrt{2}\bigl(\nu_x\sin\xi\, |X\rangle
    + \nu_y\cos\xi\, |Y\rangle\bigr)
    + \nu_z |Z\rangle.
  \end{array}\quad
\end{eqnarray}
Here the angle $\xi$ is defined by $\tan\xi =
(1-\sqrt{3})/(1+\sqrt{3})$, and $|X\rangle = |zx,xy\rangle$,
$|Y\rangle = |xy,yz\rangle$ and $|Z\rangle = |yz,zx\rangle$ are the
two-electron basis introduced in Ref.~\onlinecite{ot04}. Note that
since $\tau_z$ is diagonal in the chiral basis $|\pm \rangle$,
eigenstates of $\tau_y$ are composed of real orbitals.

\begin{figure}
\centering
\includegraphics[width=0.9\columnwidth]{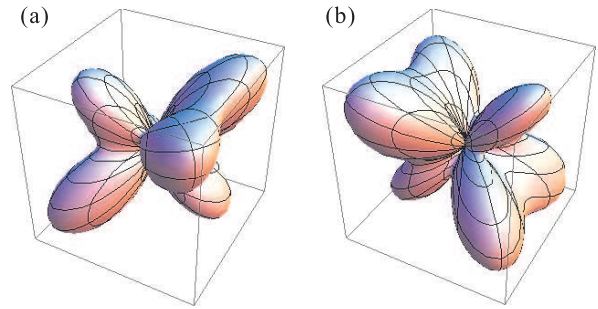}
\caption{\label{fig:rho} Electron density of state $|\tau_y =+
1\rangle$ with trigonal axis along (a) $\hat{\bm\nu}_0 = [111]$ and
(b) $\hat{\bm\nu_3} = [\bar 1\bar 11]$ directions. The explicit
forms of the density corresponding to (a) and (b) are
$\rho_0(\mathbf r) \pm \delta\rho(\mathbf r)$, respectively.}
\end{figure}

We now consider orbital chains running along $[110]$ direction, in
which the local $C_3$ axis alternates between $\hat{\bm\nu}_0$ and
$\hat{\bm\nu}_3$ [Fig.~\ref{fig:gs}(c)]. Along the chain, the
electrons are in the $|\tau_y = +1\rangle$ state whose electron
density can be readily computed
\begin{eqnarray}
    \rho(\mathbf r) &=& \frac{1}{2} \sum_{i=1,2}\int
    \delta(\mathbf r - \mathbf r_i)
    \bigl|\langle\mathbf r_1, \mathbf r_2|\tau_y = +1\rangle\bigr|^2 d^3 r_1 d^3
    r_2 \nonumber \\
    & = & \rho_0(\mathbf r) \pm \delta\rho(\mathbf r),
\end{eqnarray}
where the $+$ and $-$ signs refer to sites with $\hat{\bm\nu}_0$ and
$\hat{\bm\nu}_3$ trigonal axes, respectively. Introducing basis
functions, e.g. $\psi_{xy}(\mathbf r) = f(r)\, x y$, where $f(r)$ is
a spherically symmetric function, the uniform and staggered parts of
electron density are given by
\begin{eqnarray}
    & &\rho_0(\mathbf r) =
    \frac{1}{3} \,\psi_{xy}^2(\mathbf r) -
    \frac{2}{3}\cos\xi\sin\xi\,\psi_{yz}(\mathbf r)\,\psi_{zx}(\mathbf r)
    \\
    & & \quad + \Bigl(\frac{1}{2}-\frac{1}{3}\cos^2\xi\Bigr)\,\psi_{yz}^2(\mathbf r) +
    \Bigl(\frac{1}{2}-\frac{1}{3}\sin^2\xi\Bigr)\,\psi_{zx}^2(\mathbf r), \nonumber
\end{eqnarray}
\begin{eqnarray}
    \delta\rho(\mathbf r) = \frac{-1}{3 \sqrt{2}}\,\psi_{xy}(\mathbf
    r)\,\Bigl(\sin\xi\,\psi_{yz}(\mathbf r) +
    \cos\xi\,\psi_{zx}(\mathbf r)\Bigr).
\end{eqnarray}
The resulting density functions are plotted in Fig.~\ref{fig:rho}. A
similar density modulation, in which orbitals rotate alternatively
by about 45$^\circ$ along the orbital chain, is also observed in
first-principle density functional calculations. \cite{sarkar09}

The periodic modulation of the electron density functions along
orbital chains is a natural consequence of the staggering of local
symmetry axes. In contrast, such density-modulation is absent in
other orbital orders proposed for vanadium spinels. In $A$-type
order, the two orbital chains are characterized by occupied
two-electron states $|X\rangle$ and $|Y\rangle$, respectively. Along
a given chain, e.g. the $|X\rangle$ chain, the electron density
$\rho(\mathbf r) = \frac{1}{2}\psi_{zx}^2(\mathbf r) + \frac{1}{2}
\psi_{xy}^2(\mathbf r)$ is invariant. In the ferro-orbital order
proposed as the ground state of ZnV$_2$O$_4$, \cite{ot04} there is
only one type of orbital chain, along which the two electrons occupy
states $\frac{1}{\sqrt{2}}\bigl(|X\rangle \pm i |Y \rangle\bigr)$
alternatively, giving rise to a staggered orbital angular momentum
$\mathbf L = \pm\hat\mathbf z$. Despite the $\pi$-phase modulation,
the electron density $\rho(\mathbf r) = \frac{1}{2}
\psi_{xy}^2(\mathbf r) + \frac{1}{4}\bigl(\psi_{yz}^2(\mathbf r) +
\psi_{zx}^2(\mathbf r)\bigr)$ is the same at all sites.

\section{Magnetic order}

\label{sec:m-order}

In the absence of orbital order, interaction between vanadium spins
is governed by an isotropic Heisenberg model, which is know to
exhibit strong geometrical frustration on pyrochlore lattice.
\cite{moessner98} The energy minimum of the model is attained by a
macroscopically large number of states in which the total spin of
every tetrahedron is zero $\mathbf S_{\boxtimes} = 0$. The magnetic
frustration is partially relieved below $T_F \approx 56$ K as the
antiferromagnetic Mn-V exchange induces a ferrimagnetic order with
antialigned Mn and V spins pointing along the crystal $c$ axis. The
transverse components of V spins remain disordered.

The residual frustration is relieved by anisotropic spin exchange as
well as single-ion anisotropy in the presence of long-range orbital
order. The anisotropic spin exchange comes from the dependence of
magnetic interaction on the underlying orbital configurations, as
indicated by SE Hamiltonian~(\ref{eq:SE}). For example, orbital
order corresponding to $\mathbf l_{xy} = \pm\hat\mathbf y$ gives
rise to an anisotropic exchange constant such that $J_{[110]}
\approx \frac{4}{9} J_2$ for bonds along $[110]$ and $[1\bar 10]$
directions, i.e. directions of orbital chains, and $J_{[011]} =
J_{[010]} \approx \frac{13}{36} J_2$ for bonds along other
directions. Consequently, upon decreasing the temperature,
long-range antiferromagnetic spin correlation first develops along
orbital chains. However, three-dimensional magnetic order is not
realized due to frustrated inter-chain couplings.

\begin{figure}
\centering
\includegraphics[width=0.7\columnwidth]{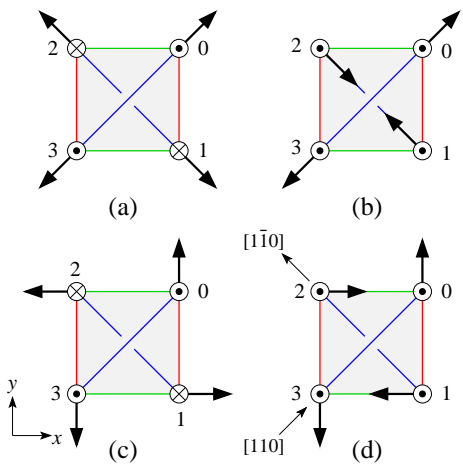}
\caption{\label{fig:m-order} Magnetic orders of vanadium spinels.
The $\odot$ and $\otimes$ symbols denote $\pm S_z$ components,
respectively. In the limit $\Delta \gg \lambda \gg J_{\tau}$, the spin
anisotropy is dominated by Eq. (\ref{eq:LS}), and the in-plane spins
point along diagonal directions $S_x = \pm S_y$. The corresponding
magnetic orders are shown in (a) and (b) for the case of
non-magnetic and magnetic $A$-site ions, respectively. In the
opposite limit $J_{\tau} \gg \Delta \gg \lambda$, spin anisotropy is
governed by Eq. (\ref{eq:LS2}), the resulting magnetic orders are
shown in (c) and (d) for the respective case of non-magnetic and
magnetic $A$-site ions.}
\end{figure}

On the other hand, single-ion anisotropy resulting from the
relativistic spin-orbit (SO) interaction $V_{LS} = \lambda (\mathbf
L\cdot\mathbf S)$ is more efficient in relieving the magnetic
frustration. It is interesting to note that the chiral basis
$|e^\pm_g\rangle$ introduced in Eq.~(\ref{eq:eg}) are simultaneous
eigenstates of angular momentum operator projected onto the local
trigonal axis:
\begin{eqnarray}
    (\mathbf L\cdot\hat{\bm\nu})\,|e^\pm_g\rangle = \pm
    |e^\pm_g\rangle.
\end{eqnarray}
Restricted to the doublet subspace $|\pm \rangle$, the angular
moment operator is given by $\mathbf L = \hat{\bm\nu}\,\tau_z$, and
the effective SO interaction becomes
\begin{eqnarray}
    \label{eq:LS}
    V_{LS} = \lambda(\mathbf S\cdot\hat{\bm\nu})\,\tau_z.
\end{eqnarray}
We first consider limit $\Delta \gg \lambda \gg J_{\tau}$, where the
single-ion physics dominates the Hamiltonian. The atomic ground
state is a non-Kramers doublet ($\lambda > 0$ in V$^{3+}$ ion)
\begin{eqnarray}
    \begin{array}{l}
    |\!\uparrow \rangle = |\tau_z = + 1\rangle \otimes
    |\mathbf S \cdot\hat{\bm\nu} = - 1\rangle, \\
    |\!\downarrow \rangle = |\tau_z = - 1\rangle \otimes
    |\mathbf S \cdot\hat{\bm\nu} = + 1\rangle,
    \end{array}
\end{eqnarray}
Long-range ordering of spins and orbitals depends further on the
relative strength of antiferromagnetic Mn-V and V-V exchanges.

In the case of a non-magnetic $A$-site ion, the ground state
consists of a uniform occupation of either $|\!\uparrow\rangle$ or
$|\!\downarrow\rangle$ states. It is easy to check that the
resulting non-collinear spin configuration ($\mathbf S_i =
+\hat{\bm\nu}$ or $\mathbf S_i = -\hat{\bm\nu}$) also minimizes the
V-V exchange. The corresponding ferro-orbital order is characterized
by order parameter $\mathbf m_{\tau} \equiv \frac{1}{4}(\bm\tau_0 +
\bm\tau_1 + \bm\tau_2 + \bm\tau_3) = \mp \hat \mathbf z$. On the
other hand, the Mn-V exchange is minimized by a staggering of
occupied $|\!\!\uparrow\rangle$ and $|\!\downarrow\rangle$ states
such that the $S_z$ component at every site is opposite to Mn spins
(whose effect can be thought of as an external magnetic field). The
ground state can then be viewed as a collection of two inequivalent
spin-orbital chains running along $[110]$ and $[1\bar 1 0]$
directions [Fig.~\ref{fig:m-order}(b)]. The corresponding orbital
configuration is described by order parameter $\mathbf l_{xy} = \pm
\hat\mathbf z$.

When $J_{\tau}$ is comparable or larger than $\lambda$, the
competition between SO interaction and orbital-exchange gives rise
to a $\tau_z \propto \lambda/J_{\tau}$. The
perturbation~(\ref{eq:LS}) thus is essentially of order $\lambda^2$.
To be consistent, we should take into account the second-order
perturbations of SO interaction simultaneously. A straightforward
calculation yields
\begin{eqnarray}
    \label{eq:LS2}
    V'_{LS} = \frac{\lambda^2}{3\Delta}\,\tau_x (S_x^2 + S_y^2 - 2 S_z^2)
    +\frac{\lambda^2}{\sqrt{3}\Delta}\,\tau_y (S_x^2 - S_y^2). \quad
\end{eqnarray}
This expression can be thought of as an invariant product of two
irreducible $e_g$ representations of $D_{3d}$. Now consider ground
state characterized by $\mathbf l_{xy} = +\hat\mathbf y$ in which
pseudospins $\bm\tau_i =\pm \hat\mathbf y$ along $[110]$ and $[1\bar
10]$ orbital chains, respectively. The second term in
Eq.~(\ref{eq:LS2}) thus introduces a {\em staggered} spin anisotropy
with easy axis parallel to $y$ or $x$ axes depending on $\tau_y =
+1$ or $-1$, respectively.

To determine the equilibrium spin configuration, we note that the
two competing anisotropies Eqs.~(\ref{eq:LS}) and (\ref{eq:LS2})
have a magnitude of order $\lambda^2/J_{\tau}$ and
$\lambda^2/\Delta$, respectively. Consequently, when orbital
exchange (including phonon-mediated exchange) dominates the trigonal
splitting $J_{\tau} \gg \Delta$, the anisotropy $\pm (S_x^2 -
S_y^2)$ wins and aligns spins to either $x$ or $y$ directions along
the respective orbital chains. The resulting magnetic orders are
shown in Figs. \ref{fig:m-order}(c) and (d) for the case of
non-magnetic and magnetic $A$-site ions, respectively. In
particular, the one shown in Fig. \ref{fig:m-order}(d) is consistent
with the proposed magnetic ground state for MnV$_2$O$_4$ in Ref.
\onlinecite{garlea08}.

\begin{figure}
\centering
\includegraphics[width=0.95\columnwidth]{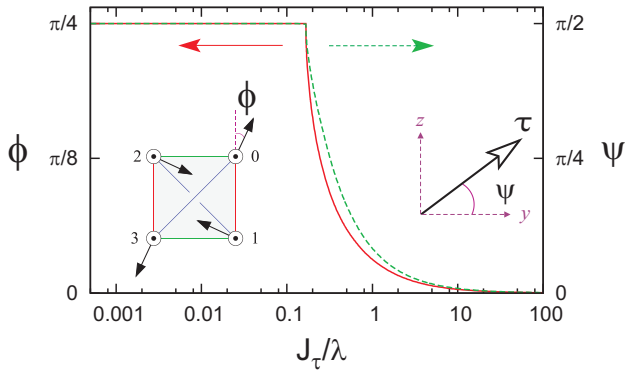}
\caption{\label{fig:phi} Evolution of orthogonal magnetic order as a
function of the ratio $J_{\tau}/\lambda$. The parameter $\psi$ and
$\phi$ measure the rotation angle of pseudospin and in-plane
orthogonal structure, respectively. The SO coupling constant is set
to $\lambda = 0.75 \Delta$, and the ferromagnetic spin component is
$S_z = 1/\sqrt{3}$. The limiting cases $\phi = \pi/4$ and $\phi = 0$
correspond to magnetic orders shown in Figs. \ref{fig:m-order} (b)
and (d), respectively.}
\end{figure}

In order to understand in more detail the transition between these
two limiting cases, we performed an explicit calculation of the
magnetic structure using Eqs. (\ref{eq:H120}), (\ref{eq:LS}), and
(\ref{eq:LS2}). Since a detailed knowledge of Mn-Mn and Mn-V
exchanges is required in order to compute the ferromagnetically
ordered $S_z$ component, we set $S_z = 1/\sqrt{3}$ to simplify the
calculation. As Fig.~\ref{fig:phi} shows, below a critical
$J^*_{\tau} \approx 0.18\lambda$, pseudospins $\bm\tau$ are
polarized along $z$ direction, while the transverse spin components
pointing along the diagonal directions form the orthogonal structure
shown in Fig.~\ref{fig:m-order}(b). Above the critical $J^*_{\tau}$,
the transverse spins rotate uniformly (the rotation is described by
angle $\phi$) while maintaining the orthogonal structure. At the
same time, pseudospins develop a finite antiferro-orbital order
along $\tau_y$ which is characterized by angle $\psi$. The
calculation shows that at $J_{\tau} \gtrsim 10 \lambda$, the spin
anisotropy is already dominated by Eq.~(\ref{eq:LS2}) as the angle
$\phi \approx 0$ and the in-plane spins essentially point along
either $x$ or $y$ axis.

\section{Conclusions}

\label{sec:conclusion}

To summarize, we have proposed and studied a spin-orbital model for
vanadium spinel MnV$_2$O$_4$ taking into account a large trigonal
distortion at the vanadium sites. Instead of conventional $t_{2g}$
triplet, our starting point is the doubly degenerate $e_g$
eigenstates of the trigonal crystal field. By introducing a
pseudospin-1/2 for the low-energy doublet, we have shown  that the
effective orbital interaction resulting from both the superexchange
and cooperative Jahn-Teller effect is described by a quantum
120$^\circ$ Hamiltonian on pyrochlore lattice. From both analytical
and numerical calculations, we have found six classical ground
states with collinear pseudospins perpendicular to either one of the
three unit vectors characterizing the anisotropic interactions. The
classical ground state is further shown to be stable against quantum
fluctuations.

The ground-state structure obtained from our model is consistent
with main experimental observations and {\em ab initio} calculation
of MnV$_2$O$_4$, namely, an antiferro-orbital order with tetragonal
$I4_1/a$ space group, a density-modulation along orbital chains, and
an orthogonal magnetic structure. The orbital order corresponding to
the semiclassical ground states consists of two inequivalent orbital
chains running along $\langle 110\rangle$ and $\langle 1\bar 1
0\rangle$ directions, similar to the so-called $A$-type
antiferro-orbital order. However, the staggering of trigonal axes
along orbital chains gives rise to a periodic variation of the
electron density function, which is absent in the $A$-type order.
Moreover, since the trigonal distortion breaks the approximate
rotational symmetry of $t_{2g}$ orbitals, orthogonal magnetic
structure is shown to be stabilized by the staggering of the
single-ion spin anisotropies.

The overall orientation of the orthogonal structure actually depends
on the relative strength of effective orbital exchange $J_{\tau}$
and spin-orbit coupling $\lambda$. The experimentally proposed
orthogonal structure \cite{garlea08} with transverse vanadium spins
pointing along either $x$ or $y$ axes is stabilized when $J_{\tau} >
\lambda$.  Using the large-$J$ approach which assumes a dominant
$\lambda$ over superexchange energy scale, we have recently shown
that the same orthogonal spin structure is stabilized when the
predominant lattice distortion is of $F_{1g}$ symmetry. In fact,
without any Jahn-Teller distortion, the large-$J$ ground state has a
collinear antiferromagnetic order similar to the one proposed in
Ref.~\onlinecite{ot04}. Noting that $J_{\tau}$ also includes
contributions from $F_{1g}$ phonons, results from the two
complementary approaches (large trigonal field vs large-$J$) are
actually consistent: stabilization of the experimentally proposed
orthogonal structure requires a dominant $F_{1g}$ distortion as well
as a smaller spin-orbit coupling. This conclusion is also supported
by recent {\em ab initio} calculation which shows that inclusion of
spin-orbit coupling does not significantly change the spin-orbital
order (a finite $LS$ coupling, however, is still required to provide
the spin anisotropies). Detailed analysis of Jahn-Teller phonons
from first-principles calculation might help clarify the role of
$F_{1g}$ phonons.

\begin{acknowledgments}
The authors acknowledge G.~Jackeli, V.~Garlea, I.~Rousochatzakis,
O.~Tchernyshyov, and R.~Valent\'i for useful discussions. G.W.C. is
particularly grateful to O.~Tchernyshyov for sharing his insights
and opinions on developing theoretical models of vanadium spinels.
\end{acknowledgments}

\appendix

\section{Phonon-mediated orbital exchange}

\label{sec:phonon}

In this Appendix, we derive the effective pseudospin interaction due
to cooperative Jahn-Teller effect. The derivation presented here is
based on a related work in Ref.~\onlinecite{chern08}. To explicitly
take into account the cooperative nature of phonon-mediated orbital
exchange, we consider coupling of orbitals to nonlocal lattice
vibrations. In fact, a similar study based on local Einstein-like
phonons gives an inconclusive result. \cite{chern08} To further
simplify the discussion, we notice that all experimentally observed
structural distortion in vanadium spinels preserves the lattice
translational symmetry. We thus restrict our analysis to phonons
with wavevector $\mathbf q = 0$.

\begin{figure}
\centering
\includegraphics[width=0.75\columnwidth]{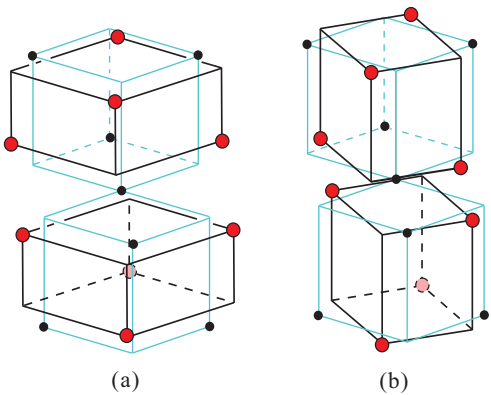}
\caption{\label{fig:phonon} $\mathbf q= 0$ lattice phonons with (a)
$E_g$ and (b) $F_{1g}$ symmetries in a spinel structure. These
normal modes are dominated by oxygen displacements. The red and
black circles indicate oxygen and vanadium ions, respectively.}
\end{figure}

Since orbitals mainly couple to oxygen ions, we focus on normal
modes which are dominated by oxygen displacements. The 8 oxygen ions
in a primitive unit cell of spinel form two tetrahedra related to
each other by inversion symmetry. In the following, we confine
ourselves to oxygen phonons with odd parity (coupling to even-parity
oxygen modes cancels identically). Among the remaining modes, we
find that the doublet $E_g$ and triplet $F_{1g}$ phonons are most
effective in JT coupling. The $E_g$ modes correspond to the
tetragonal and orthorhombic distortions of the oxygen tetrahedra,
whereas the triply degenerate $F_{1g}$ modes represent rigid
rotations of oxygen tetrahedra about the three cubic axes
(Fig.~\ref{fig:phonon}). \cite{himmrich:91}

\begin{figure}
\centering
\includegraphics[width=0.8\columnwidth]{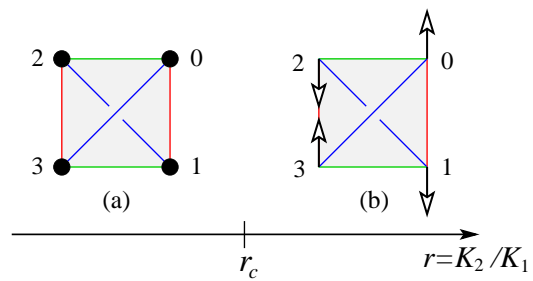}
\caption{\label{fig:jt-gs} Classical phase diagram of Hamiltonian
Eq.~(\ref{eq:cjt}). The filled circle denotes $+\tau_z$ component of
pseudospin. The ground-state orbital order depends on the ratio $r =
K_2/K_1$. The boundary separating ferro-orbital from collinear
antiferro-orbital states is given by $r_c = 8/3$.}
\end{figure}

As discussed in Sec.~\ref{sec:o-order}, the orbital doublet
$(\tau_x,\tau_y)$ of trigonal crystal field couples to the $E_g$
distortion of VO$_6$ octahedron [Eq.~(\ref{eq:jt})]. To obtain the
effective orbital interaction mediated by the above-mentioned
lattice phonons, we first express the coordinates $(\delta_1,
\delta_2)$ describing the distortion of a local VO$_6$ octahedron in
terms of coordinates of lattice $E_g$ and $F_{1g}$ phonons. The
energy costs associated with the two modes are characterized by
effective elastic constants $k_{E_g}$ and $k_{F_{1g}}$,
respectively. After integrating out the phonons, we obtain
\begin{eqnarray}
    \label{eq:cjt}
    H_{\rm JT} = -K_1 \sum_{\langle ij \rangle} \bm\tau_i\cdot\bm\tau_j
    + K_2 \sum_{\langle ij \rangle} (\bm\tau_i\cdot\hat\mathbf n_{\alpha})
    (\bm\tau_j\cdot\hat\mathbf n_{\alpha}), \quad
\end{eqnarray}
with the following effective exchange constants
\begin{eqnarray}
 K_1 = 2g^2/k_{E_g} + g^2/k_{F_{1g}}, \quad
 K_2 = 3g^2/k_{F_{1g}}.
\end{eqnarray}
The effective Hamiltonian (\ref{eq:cjt}) contains two competing
interactions: the $K_1$ term denotes an isotropic Heisenberg
exchange, whereas the $K_2$ term represents the anisotropic
120$^\circ$ interaction introduced in Sec.~\ref{sec:m120}. A
classical phase diagram of the above model is summarized in
Fig.~\ref{fig:jt-gs}, where a ferro-orbital order is separated from
the collinear antiferro-orbital ground state of the 120$^\circ$
model discussed previously. Note that the Heisenberg term has a
ferromagnetic sign, hence favoring a ferro-orbital ordering. The
ferro-orbital state is doubly degenerate with all pseudospins
pointing along either $+\hat\mathbf z$ or $-\hat\mathbf z$
directions. On the other hand, psedospins in the antiferro-orbital
phase are in a collinear up-up-down-down configuration. As discussed
in Sec.~\ref{sec:m120}, there are totally six degenerate
antiferro-orbital states; the one shown in Fig.~\ref{fig:jt-gs}(b)
is characterized by a nonzero order parameter $\mathbf l_{xy} =
+\hat\mathbf y$. By comparing the energies, we find a phase boundary
at $r_c = (K_2/K_1)_c = 8/3$.

\section{Semiclassical approach to 120$^\circ$ model}

\label{sec:semi}

Here we examine quantum corrections to the classical ground state of
orbital 120$^\circ$ model on pyrochlore lattice. Our approach is
based on a semiclassical Holstein-Primakoff expansion around the
collinear state described by order parameter $\mathbf l_{xy} =
+\hat\mathbf y$ [Fig. \ref{fig:gs}(c)]. To this end, we first
generalize the 120$^\circ$ model to pseudospins $\mathbf T$ of
arbitrary length $|\mathbf T| = \sqrt{T(T+1)}$:
\begin{eqnarray}
    \label{eq:H120T}
    H_{120^\circ} = J_{T}\sum_{\langle ij \rangle}
    (\mathbf T_i\cdot\hat\mathbf n_{\alpha})
    (\mathbf T_j\cdot\hat\mathbf n_{\alpha}),
\end{eqnarray}
where $J_T$ is an effective coupling constant. We then expand the
Hamiltonian in powers of $1/T$ around the classical ground state
using Holstein-Primakoff transformation. To simplify the
calculation, we rotate the pseudospins around $T_x$ axis such that
\begin{eqnarray}
    T_x = \tilde T_{x},\quad T_y = \pm\tilde T_{z}, \quad
    T_z = \mp \tilde T_{y},
\end{eqnarray}
where $+$ and $-$ signs refer to pseudospins along $[110]$ and
$[1\bar 10]$ chains, respectively. The classical ground state in
terms of rotated pseudospins is simply $ \tilde{\mathbf T}_i =
+T\hat\mathbf z$. We then apply the standard Holstein-Primakoff
transformation:
\begin{eqnarray}
    \label{eq:hp}
    \tilde T_z &=& T - a^\dagger a, \nonumber \\
    \tilde T_+ &=& \sqrt{2T - a^\dagger a}\,a \approx
    \sqrt{2T} \,a \\
    \tilde T_- &=& a^\dagger \sqrt{2T-a^\dagger a} \approx
    \sqrt{2T}\,a^\dagger, \nonumber
\end{eqnarray}
where $a$ and $a^\dagger$ satisfy the canonical boson commutation
relations. Substituting Eq.~(\ref{eq:hp}) into 120$^\circ$
Hamiltonian~(\ref{eq:H120}), we obtain $H_{120^\circ} = -6N_t J_{T}
T^2 + H_2 + \cdots$, where $N_t$ is the number of unit cells, $H_2$
of order $\mathcal{O}(T)$ is the quadratic magnon Hamiltonian, and
the omitted terms are of higher orders in $1/T$. We note that the
linear term $H_1 \sim \mathcal{O}(T^{3/2})$ vanishes identically as
expected for expansion around a classical ground state.

\begin{figure}
\centering
\includegraphics[width=0.95\columnwidth]{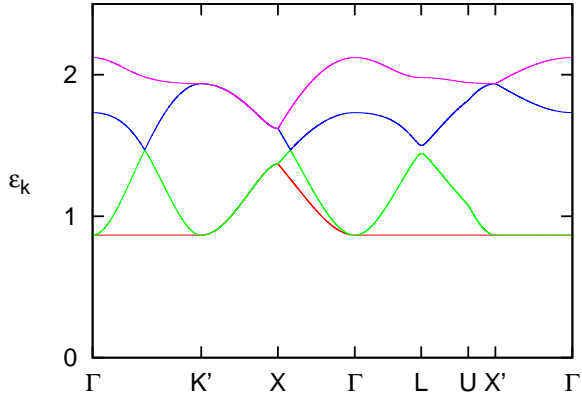}
\caption{\label{fig:spec} Quasiparticle dispersions of the
120$^\circ$-model. The energy $\varepsilon_{\mathbf k}$ is measured
in unit of $J_{T}$. The various symmetry points in $k$-space are
$\Gamma = (0,0,0)$, $K'=(1,1,0)$, $X=(1,0,0)$, $L =
(\frac{1}{2},\frac{1}{2},\frac{1}{2})$,
$U=(\frac{1}{4},\frac{1}{4},1)$, and $X'=(0,0,1)$.}
\end{figure}

After Fourier transformation, the quadratic Hamiltonian reads
\begin{eqnarray}
    H_2 = \sum_{\mathbf k} \vec a^\dagger_{\mathbf k}\,
    \left(\begin{array}{cc}
    \frac{3\tau}{2}\mathcal{I}+M_{\mathbf k} & M_{\mathbf k} \\
    M_{\mathbf k} & \frac{3\tau}{2}\mathcal{I}+ M_{\mathbf k}
    \end{array}\right) \vec a_{\mathbf k},
\end{eqnarray}
where $\vec a_{\mathbf k} = [\,a_0(\mathbf k), \cdots,a_3(\mathbf
k),a^\dagger_0(-\mathbf k),\cdots,a^\dagger_3(-\mathbf k)]^T $ is an
8-component column vector, $\mathcal{I}$ is a $4\times 4$ identity
matrix, and
\begin{eqnarray}
    M_{\mathbf k} = \frac{T}{8}\left(\begin{array}{cccc}
    0 & c_{yz} & c_{zx} & 4 c_{xy} \\
    c_{yz} & 0 & 4 \bar c_{x y} & \bar c_{z x} \\
    c_{zx} & 4 \bar c_{x y} & 0 & \bar c_{y z} \\
    4 c_{xy} & \bar c_{z x} & \bar c_{y z} & 0
    \end{array}\right). \\ \nonumber
\end{eqnarray}
For convenience, we have defined $c_{\alpha\beta} =
\cos\bigl[(k_{\alpha} + k_{\beta})/4\bigr]$ and $\bar c_{\alpha
\beta} = \cos\bigl[(k_{\alpha}-k_{\beta})/4\bigr]$ (the lattice
constant is set to 1). To diagonalize $H_2$, we consider equation of
motion for boson operator $\vec a_{\mathbf k}$:
\begin{eqnarray}
    i\frac{\partial\vec a_{\mathbf k}}{\partial t}=
    \left(\begin{array}{cc}
    \frac{3\tau}{2}\mathcal{I}+M_{\mathbf k} & M_{\mathbf k} \\
    -M_{\mathbf k} & -\frac{3\tau}{2}\mathcal{I}- M_{\mathbf k}
    \end{array}\right)\,\vec a_{\mathbf k} \equiv
    \mathcal{L}_{\mathbf k}\,\vec a_{\mathbf k}. \quad
\end{eqnarray}
From eigenvectors of matrix $\mathcal{L}_{\mathbf k}$, one can
construct a canonical transformation $\vec a_{\mathbf k} =
\mathcal{T}_\mathbf k\,\vec c_{\mathbf k}$, where matrix
$\mathcal{T}_{\mathbf k}$ satisfies
\begin{eqnarray}
    \label{eq:Tx}
    \mathcal{T}^\dagger_{\mathbf k} \eta \mathcal{T}_{\mathbf k} =
    \eta, \quad\quad
    \mathcal{T}^{-1}_{\mathbf k} \mathcal{L}_{\mathbf k}
    \mathcal{T}_{\mathbf k} = \eta\Lambda_{\mathbf k}.
\end{eqnarray}
Here the diagonal matrices $\eta = {\rm
diag}(\mathcal{I},-\mathcal{I})$ and $\Lambda_{\mathbf k} = {\rm
diag}\bigl(\varepsilon_0(\mathbf k), \cdots, \varepsilon_3(\mathbf
k), \varepsilon_0(-\mathbf k), \cdots,\varepsilon_3(-\mathbf
k)\bigr)$. Using Eq.~(\ref{eq:Tx}), the diagonalized Hamiltonian
becomes
\begin{eqnarray}
    H_2 = \frac{1}{2}\sum_{\mathbf k} \vec c^\dagger_{\mathbf k}
    \Lambda_{\mathbf k} \vec c_{\mathbf k} = \sum_m \sum_{\mathbf k}
    \varepsilon_m(\mathbf k) c^\dagger_m(\mathbf k) c_m(\mathbf k).
    \quad
\end{eqnarray}
By setting $T$ to its physical value $1/2$, the numerically obtained
dispersion $\varepsilon_m(\mathbf k)$ along various symmetry
directions of the Brillouin zone is shown in Fig.~\ref{fig:spec}.
The spectrum is fully gapped with an energy gap of order $J_{T}$,
implying a small quantum correction. To confirm this, we also
compute the harmonic correction to the sublattice `magnetization'
$\langle \tilde T_z \rangle = 1/2 - \langle a^\dagger_i a_i
\rangle$. Using the explicit expression of the canonical
transformation $\mathcal{T}_{\mathbf k}$:
\begin{eqnarray}
    a_s(\mathbf k) = u_{s,m}(\mathbf k)\, c_m(\mathbf k)
    + v_{s,m}(\mathbf k)\, c^\dagger_m(-\mathbf k),
\end{eqnarray}
the average quasiparticle number can be computed: $\langle
a^\dagger_i a_i \rangle = (1/N_t)\sum_{\mathbf k} \langle
a^\dagger_s(\mathbf k) a_s(\mathbf k) \rangle = \sum_m\sum_{\mathbf
k} v_{s,m}^2$. The numerically obtained sublattice `magnetization'
is given by $\langle \tilde T_z \rangle \approx 1/2 - 0.02154$. The
collinear ground states shown in Fig. \ref{fig:gs} thus are stable
against quantum fluctuations, at least at the harmonic order.


\begin{thebibliography}{99}


\bibitem{pen97} H. F. Pen, J. ven den Brink, D. I. Khomskii, and G.
A. Sawatzky, Phys. Rev. Lett. {\bf 78}, 1323 (1997).

\bibitem{radaelli02} P.G. Radaelli, Y. Horibe, M. J. Gutmann,
H. Ishibashi, C. H. Chen, R. M. Ibberson, Y. Koyama, Y.S. Hor, V.
Kiryukhin, and S. W. Cheong, Nature (London) {\bf 416}, 155 (2002).

\bibitem{schmidt04} M. Schmidt, W. Ratcliff, II, P. G. Radaelli, K. Refson, N. M. Harrison, and S. W.
Cheong, Phys. Rev. Lett. {\bf 92}, 056402 (2004).

\bibitem{horibe06} Y. Horibe, M. Shingu, K. Kurushima, H. Ishibashi,
N. Ikeda, K. Kato, Y. Motome, N. Furukawa, S. Mori, and T.
Katsufuji, Phys. Rev. Lett. {\bf 96}, 086406 (2006).



\bibitem{dimatteo04} S. Di Matteo, G. Jackeli, C. Lacroix, and N. B.
Perkins, Phys. Rev. Lett. {\bf 93}, 077208 (2004).

\bibitem{khomskii05} D. I. Khomskii and T. Mizokawa, Phys. Rev.
Lett. {\bf 94}, 156402 (2005).

\bibitem{moessner98} R. Moessner and J. T. Chalker,
Phys. Rev. Lett. {\bf 80}, 2929 (1998).


\bibitem{mamiya97} H. Mamiya, M. Onoda, T. Furubayashi, J. Tang, and
I. Nakatani, J. Appl. Phys. {\bf 81}, 5289 (1997).


\bibitem{reehuis03} M. Reehuis, A. Krimmel, N. B\'uttgen, A. Loidl, A.
Prokofiev, Eur. Phys. J. B {\bf 35}, 311 (2003).

\bibitem{onoda03} M. Onoda and J. Hasegawa,
J. Phys.: Condens. Matter {\bf 15}, 95 (2003).

\bibitem{lee04} S.-H. Lee, D. Louca, H. Ueda, S. Park, T. J. Sato, M.
Isobe, Y. Ueda, S. Rosenkranz, P. Zschack, J. Iniguez, Y. Qiu, R.
Osborn, Phys. Rev. Lett. {\bf 93}, 15640 (2004).


\bibitem{adachi05} K. Adachi, T. Suzuki, K. Kato, K. Osaka, M.
Takata, and T. Katsufuji, Phys. Rev. Lett. {\bf 95}, 197202 (2005).

\bibitem{suzuki07} T. Suzuki, M. Katsumura, K. Taniguchi, T. Arima,
and T. Katsufuji, Phys. Rev. Lett. {\bf 98}, 127203 (2007).


\bibitem{garlea08} V. O. Garlea, R. Jin, D. Mandrus, B. Roessli,
Q. Huang, M. Miller, A. J. Schultz, and S. E. Nagler, Phys. Rev.
Lett. {\bf 100}, 066404 (2008).

\bibitem{chung08} J.-H. Chung, J.-H. Kim, S.-H. Lee, T. J. Sato, T. Suzuki, M. Katsumura, and T.
Katsufuji, Phys Rev. B {\bf 77}, 054412 (2008).


\bibitem{tsunetsugu03} H. Tsunetsugu and Y. Motome, Phys. Rev. B
{\bf 68}, 060405(R) (2003).


\bibitem{ot04} O. Tchernyshyov, Phys. Rev. Lett. {\bf 93}, 157206
(2004).


\bibitem{dimatteo05} S. Di Matteo, G. Jackeli, and N. B. Perkins,
Phys. Rev. B {\bf 72}, 020408(R) (2005).


\bibitem{maitra07} T. Maitra and R. Valenti, Phys. Rev. Lett. {\bf
99}, 126401 (2007).

\bibitem{pardo08} V. Pardo, S. Blanco-Canosa, F. Rivadulla,
D.I. Khomskii, D. Baldomir, Hua Wu, and J. Rivas, Phys. Rev. Lett.
{\bf 101}, 256403 (2008).

\bibitem{chern08} G.-W. Chern, PhD thesis, Johns Hopkins University,
Baltimore (2008).

\bibitem{chern09} G.-W. Chern and N. B. Perkins, Phys. Rev. B {\bf 80}, 180409(R) (2009).


\bibitem{sarkar09} S. Sarkar, T. Maitra, R. Valent\'i, and T. Saha-Dasgupta,
Phys. Rev. Lett. {\bf 102}, 216405 (2009).

\bibitem{baek09} S.-H. Baek, N. J. Curro, K.-Y. Choi, A. P. Reyes, P. L. Kuhns,
H. D. Zhou, and C. R. Wiebe, Phys. Rev. {\bf 80}, 140406 (2009).


\bibitem{nussinov04} Z. Nussinov, M. Biskup, L. Chayes,
and J. v. d. Brink, Europhys. Lett. {\bf 67}, 990 (2004).

\bibitem{biskup05} M. Biskup, L. Chayes, and Z. Nussinov,
Commun. Math. Phys. {\bf 255} 253 (2005).


\bibitem{castellani} C. Castellani, C. R. Natoli, and J. Ranninger,
Phys. Rev. B {\bf 18}, 4945 (1978); {\bf 18}, 4967 (1978); {\bf 18},
5001 (1978).

\bibitem{mila00} F. Mila, R. Shiina, F.-C. Zhang, A. Joshi,
M. Ma, V. Anisimov, and T. M. Rice, Phys. Rev. Lett. {\bf 85}, 1714
(2000).



\bibitem{perkins0209} S. Di Matteo, N.B. Perkins and   C.R. Natoli, Phys. Rev. B {\bf 65},
054413 (2002); N.B. Perkins, S. Di Matteo and   C.R. Natoli, Phys. Rev. B {\bf 80}, 165106 (2009).

\bibitem{anisimov99} V. I. Anisimov, M. A. Korotin, M. Zolfl, T.
Pruschke, K. Le Hur, and T. M. Rice, Phys. Rev. Lett. {\bf 83}, 364
(1999).


\bibitem{brink99} J. van den Brink, P. Horsch, F. Mack, and A. M.
Oles, Phys. Rev. B {\bf 59}, 6795 (1999).


\bibitem{khomskii03} D. I. Khomskii and M. V. Mostovoy,
J. Phys. A - Math. Gen. {\bf 36} 9197 (2003).


\bibitem{note} The $e_g$ system discussed in Refs.~\onlinecite{brink99,khomskii03} refers to the degenerate
orbitals $d_{3z^2-r^2}$ and $d_{x^2-y^2}$ in ions with, e.g. $d^9$
configuration. It should not be confused with the $e_g$ doublet
(\ref{eq:eg}) derived from the $t_{2g}$ orbitals.


\bibitem{zhao08} E. Zhao and W. V. Liu, Phys. Rev. Lett. {\bf 100},
160403 (2008).

\bibitem{wu08} C. Wu, Phys. Rev. Lett. {\bf 100}, 200406
(2008).


\bibitem{kk} K. I. Kugel and D. I. Khomskii, Sov. Phys. Usp. {\bf 25}, 231
(1982).

\bibitem{nussinov05} Z. Nussinov and E. Fradkin, Phys. Rev. B, {\bf 71}, 195120
(2005); C. D. Batista and Z. Nussinov, Phys. Rev. B {\bf 72}, 045137
(2005).


\bibitem{note1} It should be noted that, contrary to the case of bipartite cubic
lattice, antiferromagnetic and ferromagnetic 120$^\circ$-models on
pyrochlore lattice are inequivalent.


\bibitem{chern06} G.-W. Chern, C. Fennie, and O. Tchernyshyov,
Phys. Rev. B {\bf 74}, 060405(R) (2006).


\bibitem{himmrich:91} J. Himmrich and H. D. Lutz, Solid State Comm.
{\bf 79}, 447 (1991).


\end{thebibliography}
\end{document}